\documentstyle[12pt,aaspp4]{article}

\lefthead{Distribution of Size and Lifetime of Bright Points}
\righthead{Abramenko et al.}

\begin{document}

\title{Statistical Distribution of Size and Lifetime of Bright Points
Observed with the New Solar Telescope}

\author{Valentyna Abramenko, Vasyl Yurchyshyn, Philip Goode, Ali Kilcik}
\affil{Big Bear Solar Observatory, 40386 N. Shore Lane, Big Bear City, CA
92314}

\begin{abstract}
We present results of two-hour non-interrupted observations of solar granulation
obtained under excellent seeing conditions with the largest aperture
ground-based solar telescope - the New Solar Telescope (NST) - of Big Bear
Solar Observatory.
Observations were performed with adaptive optics correction using a broad-band
TiO filter in the 705.7 nm spectral line with a time cadence of 10 s and a pixel
size of 0.0375''. Photospheric bright points (BPs) were detected and tracked. We
find that the BPs detected in NST images are co-spatial with those visible in
Hinode/SOT G-band images. In cases where Hinode/SOT detects one large BP, NST
detects several separated BPs. Extended filigree features are clearly fragmented
into separate BPs in NST images. The distribution function of BP sizes extends
to the diffraction limit of NST (77 km) without saturation and corresponds to a
log-normal distribution. The lifetime distribution function follows a log-normal
approximation for all BPs with lifetime exceeding 100 s. A majority of BPs are
transient events reflecting the strong dynamics of the quiet sun:  98.6\% of BPs
live less than 120 s. The longest registered life time was 44 minutes. The size
and maximum intensity of BPs were found to be proportional to their life times.

\end{abstract}


\section { Introduction}

Bright points (BPs) observed in the solar photosphere in FeI 6173 \AA\
filtergrams
(Muller et al. 2000) and in G-band images (Berger \& Title 2001; Ishikawa et al.
2007) are shown to be co-spatial and co-temporal with magnetic
elements. However, only about 20\% of magnetic elements in intranetwork  areas
are related to BPs (de Wijn et al. 2008; Ishikawa et al. 2007). Photospheric
BPs, therefore, represent a subset of the magnetic elements. The relationship
between BPs and magnetic field seems to be so solid that the former are
frequently addressed as magnetic bright points (e.g., Utz et al. 2009, 2010).
The mechanism for the formation of BPs is thought to be a convective collapse of
a magnetic flux tube (Parker 1978; Spruit 1979) implying strong downflows and
evacuation of the flux tube allowing us to look deeper and see hotter plasma.
Analysis of statistical distributions of BPs can shed a light on the fundamental
properties of smallest magnetic elements.

Another strong reason for enhanced interest in BPs is that they are very
reliable tracers of transverse motions of the footpoints of photospheric
magnetic flux elements. These motions are thought to be an ultimate source of
the energy needed to heat the chromosphere and corona, either via waves or via
magnetic reconnection of intertwined flux tubes (e.g., reviews by Cranmer
(2002), Cranmer \& van Ballegooien (2005), and Klimchuk (2006) and references
herein). These and other considerations stimulated numerous studies of
transverse (to the line-of- sight) velocities and various applications of the
results to the problems of wave generation (Cranmer \& van Ballegooien 2005) and
reconnection (Cranmer \& van Ballegooien 2010 and references therein).

Studies of BPs size and lifetime distributions, however, are rather scanty.
Based on Hinode/SOT (Kosugi et al. 2007; Tsuneta et al. 2008) observations, Utz
et al. (2009, 2010) studied statistical distributions of G-band BP's sizes and
lifetimes. They used a 5.7 hour data with a pixel size of 0.054${''}$ and
0.109${''}$, and a time cadence of about 32 s. The authors reported a nearly
Gaussian distribution of BP's size, with the peak located near the diffraction
limit of the telescope (157 km). For the lifetime distribution, an exponential
fit was found.

Now, with the New Solar Telescope (NST) at Big Bear Solar Observatory
(BBSO) operational, we attempt to reveal the BPs distribution
functions on the basis of more accurate data, which are described in detail in
the next session.

\section { Data }

Observations of solar granulation were performed with the NST (Goode et al.
2010) with a 1.6 meter clear aperture using a broadband TiO filter centered at
a wavelength of 705.7 nm. This spectral line is sensitive to temperature, and it
is usually used to observe the sunspot umbra/penumbra (Berdyugina et al. 2003;
Reithmuller et al. 2008). When observing granulation with this line, a dual
effect presents itself: for granules and BPs the intensity is the same as
observed in continuum, whereas for dark cool intergranular lanes, the observed
intensity is lowered due to absorption in the TiO line. Thus, TiO images provide
an enhanced gradient of intensity around BPs, which is very beneficial for
imaging them.

The FOV of the broadband filter imager is $76.8\times0.76.8{''}$, and the pixel
scale of the PCO.2000 camera we used is 0.0375${''}$. This pixel size is 2.9
times smaller than the Rayleigh diffraction limit of NST, $\theta_1
=1.22\lambda/D = 0.11{''} =77$ km, and 2.5 times smaller than the FWHM of the
smallest resolved
feature\footnote{http://www.telescope-optics.net/telescope-resolution.htm}
($\theta_2=1.03\lambda/D = 65$ km). 

Uninterrupted observations of a quiet sun area near the disk center were
performed on August 3, 2010 between 17:06 UT and 18:57 UT. The observations were
made with tip-tilt and adaptive optics corrections (Cao et al. 2010). The time
cadence of the speckle reconstructed images was 10 s. To obtain one speckle
reconstructed image, we took a burst of 70 recorded with 1 ms exposure time,
Then we applied the KISIP speckle reconstruction code (Woger \& von der Luhe
2008) to each burst. Resulting images were carefully aligned and de-stretched.
In the present study, we utilized only the central part (28.3$\times$26.2${''}$
of the FOV, where the adaptive optics corrections were most efficient. The
resulting data set consisting of 648 images was used to detect and trace BPs. A
movie of this data set can be found on the BBSO
website\footnote{http://bbso.njit.edu/nst-galery.html}.

On the same day, Hinode/SOT obtained a synoptic G-band image within the time
interval of our observations. In Figure 1 we show two images of the
same area on the Sun simultaneously recorded with the Hinode/SOT (left) and NST
(right). The Hinode image was processed with standard data-reduction Solar
SoftWare prep-fg.pro
package\footnote{http://msslrx.mssl.ucl.ac.uk:8080/SolarB/AnalysisSoftware.jsp}.
All BPs visible in the Hinode image are present in the NST image. In addition,
NST sees much more: small single BPs are clearly visible inside dark
intergranular lanes (examine the bottom halves of the images). Moreover, in
places where Hinode/SOT detects a single large BP, NST detects several separated
BPs (center of the image). Extended filigree features are clearly fragmented
into separate BPs in the NST image.

%#####################################################################
\begin{figure}[!ht] \centerline {
\epsfxsize=6.0truein
\epsffile{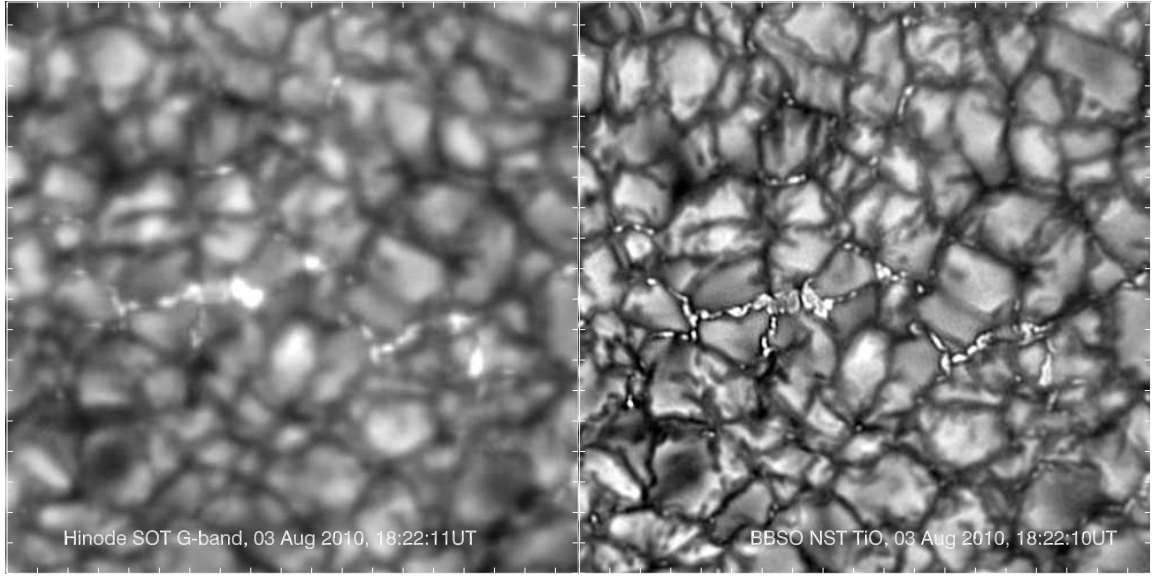}}
\caption{{\it Left - } Hinode G-band image obtained on 2010 August 3 at
18:22:11 UT (pixel size 0.109${''}$). {\it Right - }
NST TiO image obtained on
2010 August 3 at 18:22:10 UT (pixel size 0.0375${''}$). Both images cover the
same area of
18.8$\times$18.8${''}$ on the Sun.}
\label{fig1}
\end{figure}
%#####################################################################

\section { Method }

To detect and track BPs, we took advantage of the three most important
properties of BPs (Utz et al. 2009): small size, enhanced intensity and strong
gradient in intensity around BPs. We calculated a BP-mask as follows. First, we
smoothed each image (by applying a three-point running average), and then we
subtracted the smoothed image from the original image. This allowed us to
substantially suppress the intensity in granules and only slightly diminish the
peak intensity in BPs. Then we applied a thresholding technique to produce a
BP-mask. First, we chose the lowest possible value of the threshold, $th=85$
data numbers (DN), so that all BPs outside granules were selected. 
We then contoured and numerated each BP on an image. The routine was repeated
for each image. 
For comparison, we also run our code with $th=120$ DN, which
allowed us to only select brightest BPs. An example of BPs detection with
$th=85$ DN is shown in Figure 2.
%#####################################################################
\begin{figure}[!ht] \centerline {
\epsfxsize=6.0truein
\epsffile{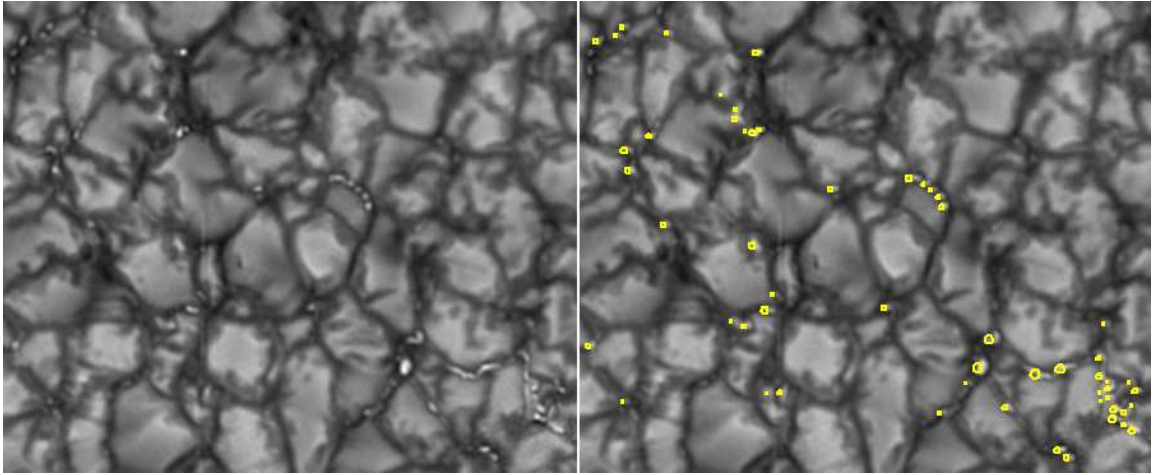}}
\caption{NST TiO image taken at 17:08:15 UT ({\it left}). The right panel
shows the same image with detected BPs denoted by yellow pixels.The image size
is                                              
16.9${''}$ $\times$ 13.9${''}$. }
\label{fig2}
\end{figure}
%#####################################################################
For each BP, we determined its area, its maximal intensity, and the equivalent
diameter, $d$, of the BP - the diameter of a circle whose area is equal to the
BP's area. Coordinates $(x_c,y_c)$ of the center of the equivalent circle were
determined as mean values of corresponding coordinates of all pixels belonging
to the BP. Centers of the circles were labeled.

We then traced the detected BPs from one image to another. When in image $i$,
inside the radius of $d/2$ around the $(x_c,y_c)$ pixel, we find a pixel labeled
as a center of a BP in the $i+1$ image, we assigned the two found objects to be
the same BP visible on two consecutive images. This approach allows us to trace
sub-sonic displacements of BPs of radius larger than 3 pixels. For smaller BPs
we cannot trace large displacements (velocities larger than 4-5 km s$^{-1}$). It
was undesirable to choose more broad than $d/2$ for the radius for searching the
response object in the next image because BPs are frequently located very close
to each other. When a BP merged with another BP, or when we could not find a
BP in four consecutive images, we concluded that this BP ceased to exist. We
thus were able to determine the lifetime of each tracked BP. The diameter, $D$, 
of each BP was determined by averaging over the lifetime equivalent diameter,
$d$. Below in this paper, by ''BP'' we mean an object observed at several
consecutive images, but not a single spot of enhanced brightness observed in a
single image.

When tracking BPs, we discarded all events with the area smaller than 2 pixels,
all events detected at only one image, as well as events with maximal intensity
below the mean intensity of the image. All these restrictions resulted in total
of $N=$13597 tracked BPs in case of $th=$85 DN threshold and $N=$7148 BPs for
$th=$120 DN.

\section { Results }

Figure 3 shows the probability distribution functions (PDFs) for the
diameter, $D$, lifetime, $LT$, and maximum intensity, $I_{max}$, of BPs. In the
left panels, results for two runs are shown: the black lines correspond to
$th=$85 DN and the red lines to $th=$120 DN. The right panels show PDFs only
from the $th=$85 DN run. For all the PDFs in the figure, the difference between
the two runs is small: for higher threshold the relative number of large-size
events slightly decreases, and the relative number of brighter events slightly
increases. The lifetime distribution does not appear to be affected by the
threshold at all. The reason for this might be the very sharp gradient of
intensity around BPs, so that the change in the cutoff level has little effect
on the size of the selected BP. This is in line with Utz et al. (2009), who
reported that a change in the cutoff level by 30\% results in change of sizes by
only about 10\%. Further, in the text, we only discuss the PDFs obtained for the
$th=$85 DN threshold.

The PDF of diameters displays a nearly linear behavior in the linear-logarithmic
plot (Figure 3, a). The monotonous increase is present down to the
diffraction limit, $\theta_1$, of NST. The saturation and the turnover of the
observed PDF at scales smaller than $\theta_1$ (to the left from the dashed
vertical line) might be caused either by the fact than the NST does not purely
resolve elements smaller than 77 km, or that elements much smaller than 77 km do
not exist. For comparison, in the same panel we show the BP size distribution as
derived from Hinode/SOT G-band images by Utz et al. (2009). In spite of a
drastic difference in the shape of the PDFs derived from the two instruments,
physically, they are consistent. The Hinode PDFs also saturate at scales
corresponding to the diffraction limit of the instrument, which is 157 km for
SOT observations in the G-band.

From these results, we may conclude that the real minimum size of magnetic BPs
(or magnetic flux tubes) has not yet been detected in observations with modern
high resolution telescopes. As to the maximum size of BPs, the upper boundary
for it seems to exist: as soon as BPs defined as those located inside
intergranular lanes, the diameter of BPs cannot be larger than characteristic
width of the intergranular lanes, which varies in a range of 150-400 km.

To determine the analytical fit for the distribution functions, PDF$(u)$, where
$u$ denotes $D, LT, I_{max}$, we applied three different approximations:
exponential, power law, and log-normal. The exponential function can be written
as
\begin{equation}
PDF(u) = exp(-\beta u + c_1),
\label{ex}
\end{equation}
where $\beta$ and $c_1$ are free parameters, while the power law approximation
can be presented as
\begin{equation}
PDF(u) = 10^{c_2} u^{\alpha}
\label{pl}
\end{equation}
with free parameters $\alpha$ and $c_2$.
The log-normal distribution function (the logarithm of $u$ is normally
distributed, e.g., see Aitchison \&
Brown 1957; Romeo et al. 2003; Abramenko \& Longcope 2005) is
\begin{equation}
PDF(u) = \frac{1}{us\sqrt{2\pi}} exp \left ( - \frac{1}{2} \left ( \frac{ln(u)-m}{s}\right )^2 \right ),
\label{ln}
\end{equation}
where $m$ and $s$ are the mean value and the standard deviation of $ln(u)$.

We applied the above approximations covering the largest possible interval,
$\Delta$, where the fit was performed with a minimum reduced $\chi^2$ test.
Results are presented in Figure 3 and the parameters of the fits are
listed in Table 1.

For the BP size distribution function, PDF$(D)$, is a log-normal fit since it has
the lowest $\chi^2$ while the fitting interval is the largest. The mode of the
log-normal distribution, $e^m$, occurs at $D = 64$ km, which is lower than the
diffraction limit. It is not excluded that observations with even higher
resolution will show the mode at even smaller scales. What is obvious now is that
the size distribution is not scale-free, since the corresponding power-law fit
is applicable over only a narrow range. Moreover, the power-law $\chi^2$ is the
worst.

The lifetime distribution function (Figure 3, c, d and Table 1, 3rd
column) can be fitted very well by all three approximations, however,
differently for different intervals. Short-lived BPs are better approximated by
the power-law fit, while medium lifetime BPs (500 - 1200 s) better obey an
exponential law. When we include long-lived BPs, the log-normal fit seems to be
the best suitable for all observed BPs that live longer than 100 s. Note that
the longest living BP in our data set was observed during 44.2 minutes, which is
less than half of the length of the data set.

The maximum intensity distribution function (Figure 3, e, f; Table 1,
the right-most column) can be equally satisfactorily fitted with an exponential
or power law with the best $\chi^2$ test favoring the power law. At the same
time, the log-normal fit is not suitable at all for PDF$(I_{max})$. Inside the
range of approximately 3800 - 4500 DN, the function is flat indicating a nearly
uniform distribution of BPs with the maximum intensities in this range.
Note that this is an interval of typical intensities found inside granules.
Recall that we discarded all BPs with $I_{max} < \langle I \rangle \approx 3400$
DN, so that the saturation of the PDF$(I_{max})$ at 3800 - 4500 DN is not related
to the BP selection routine.  The flat range of PDF$(I_{max})$ might be caused
either by insufficient sensitivity of NST to weak BPs, or by a real deficit of
weak BPs.

The regression plots of the diameter and maximum intensity of BPs versus their
lifetime (Figure 4) show the presence of a direct proportionality: BPs
that live longer, tend to be larger and brighter.

%#####################################################################
\begin{figure}[!ht] \centerline {
\epsfysize=6.2truein
\epsffile{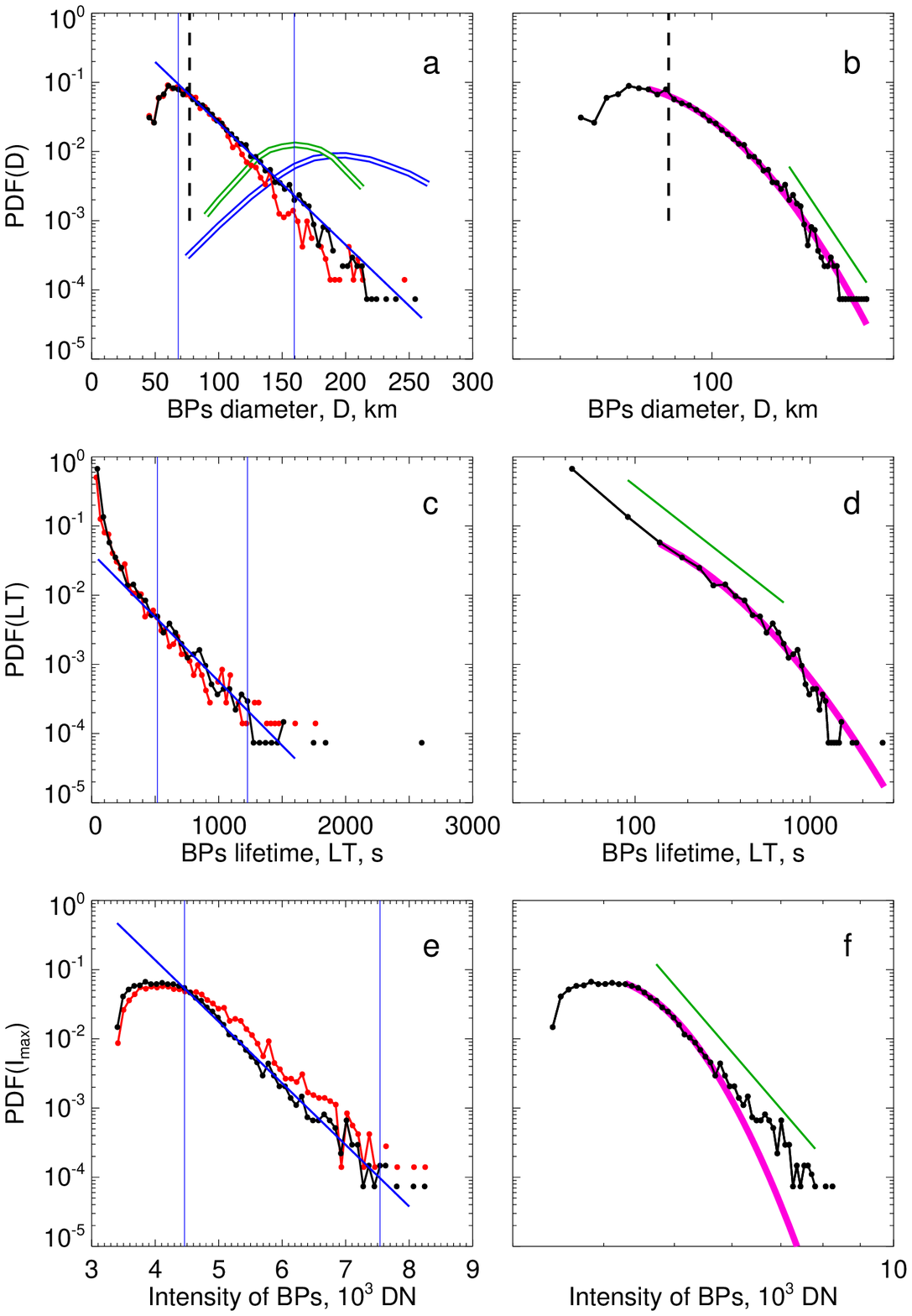}}
\caption{PDFs of the BPs diameter ({\it a,b}),lifetime  ({\it c,d}), and
maximum intensity ({\it e,f}). Left column shows the data for both runs: for the
mask threshold of 85 DN ({\it black}), and 120 DN ({\it red}). Straight thick
blue
lines in all left frames show the exponential fit to the data points calculated
inside a range between the vertical thin blue lines. In all the right frames,
the log-normal and power-law fits were applied inside
intervals, $\Delta$, covered by the purple and green lines, respectively. In the
frames {\it a,b}, the vertical dashed segment shows the position of the Rayliegh
diffraction limit of NST for observations with the TiO filter. Blue and green
double curves show the distribution of BPs size from Hinode G-band observations
as reported by Utz et al. (2009): blue (green) line for the pixel size of
0.109${''}$ (0.054${''}$).}
\label{fig3}
\end{figure}
%#####################################################################

%#####################################################################
\begin{figure}[!ht] \centerline {
\epsfxsize=3.0truein
\epsffile{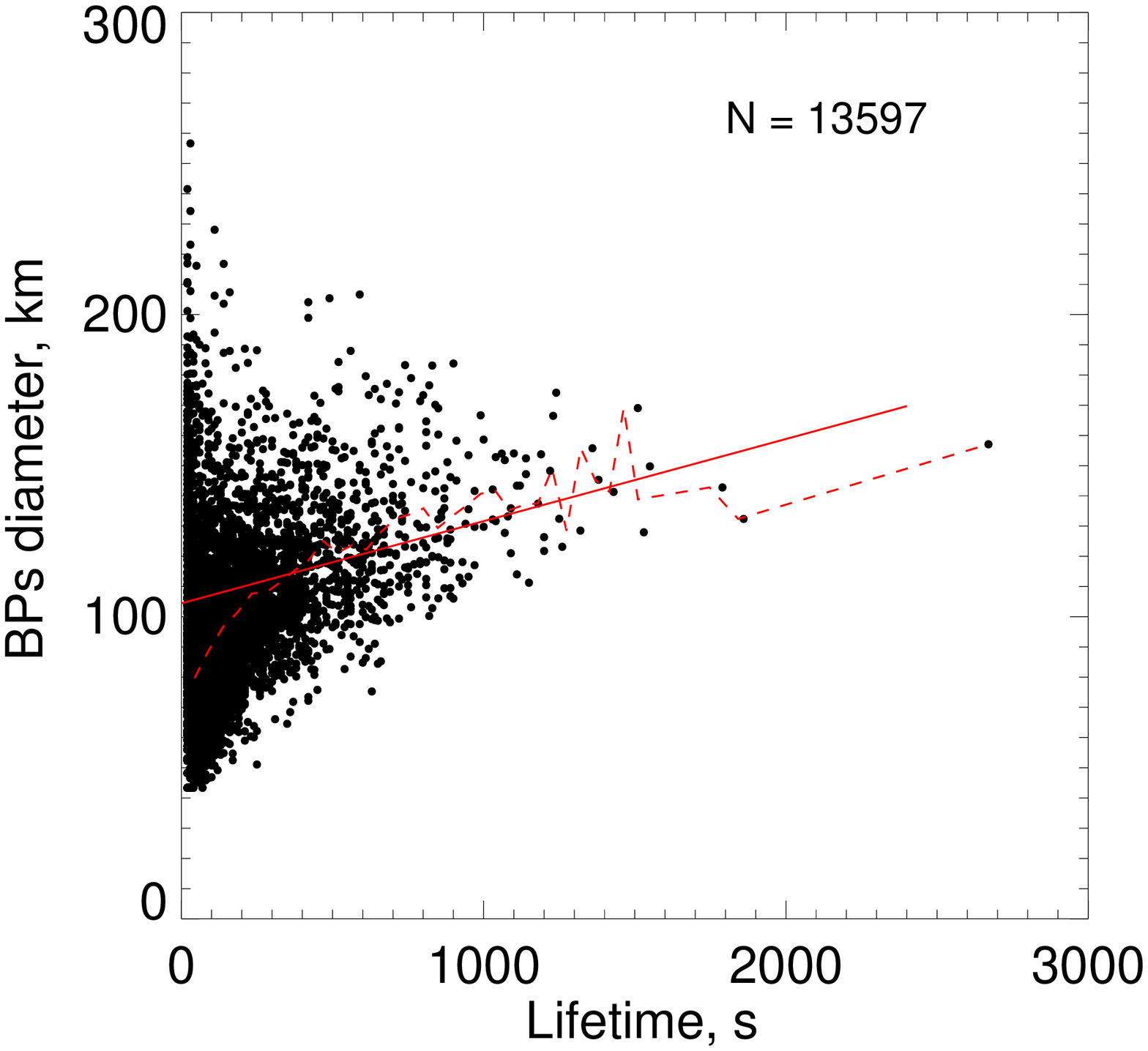}
\epsfxsize=3.0truein
\epsffile{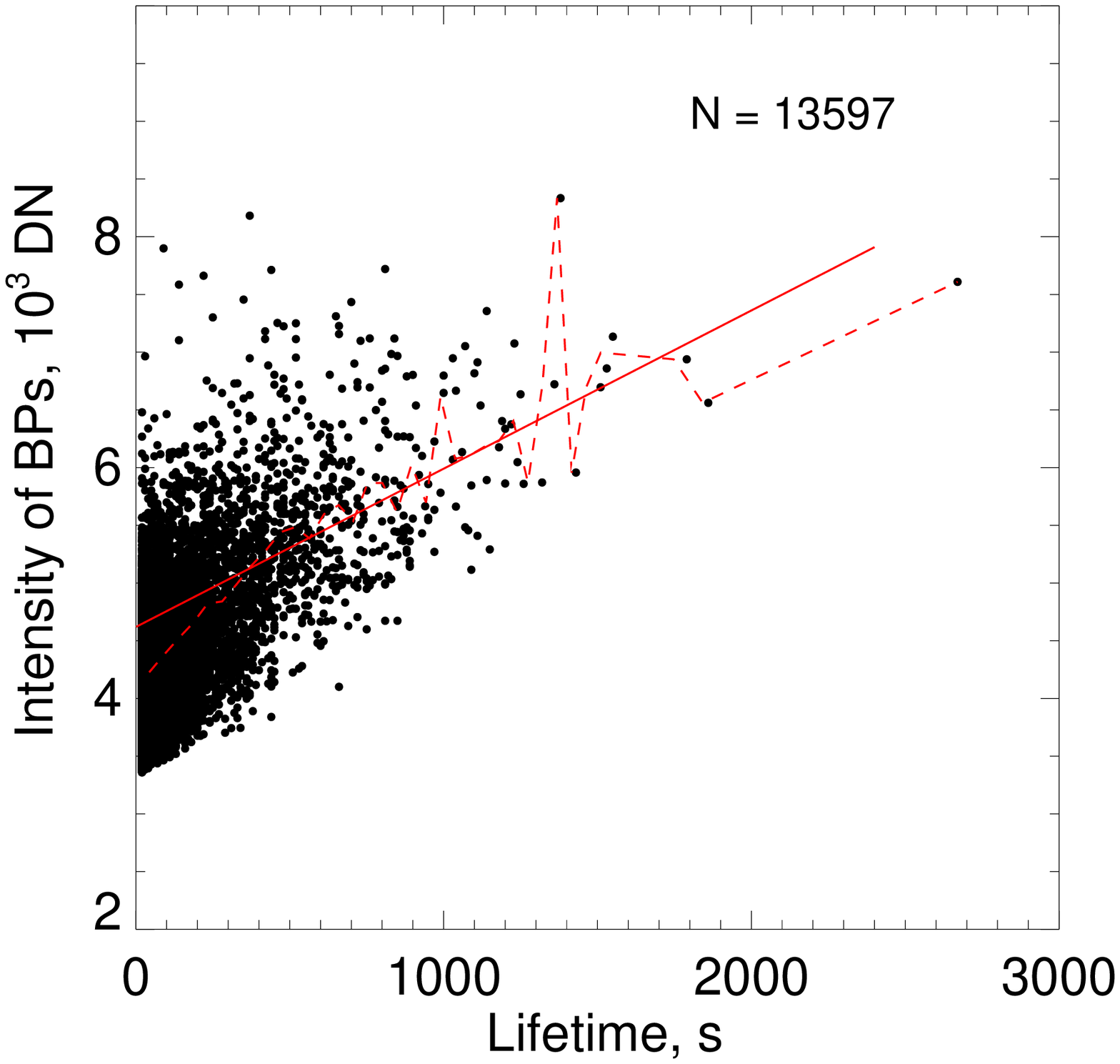}}
\caption{BPs lifetime plotted versus the BPs diameter ({\it left}) and
maximum intensity ({\it right}). Dashed lines connect the mean values
calculated inside each bin of the lifetime distribution function, PDF$(LT)$.
Straight solid lines are the best linear fit to the dashed line points.}
\label{fig4}
\end{figure}
%#####################################################################

\section { Conclusions }

A two-hour data set of quiet sun granulation obtained with highest to-date
spatial and temporal resolution (77 km diffraction limit NST images with a
time cadence of 10 s) was analyzed to obtain statistical distributions of size,
lifetime and maximum intensity of photospheric BPs. An NST TiO image was
compared with co-temporal and co-spatial image from Hinode/SOT obtained in
G-band.

NST BPs are co-spatial with those visible in Hinode/SOT G-band data.
This result allows us to take advantages of observations with the TiO filter,
in contrast to the G-band filter: the adaptive optics system and the speckle
reconstruction code are more efficient  when observing at longer wavelengths.
We see a clear improvement caused by the higher resolution of the NST: in cases
where Hinode/SOT detects one large BP, NST detects several separated BPs.
In addition, NST detects numerous small and weak BPs which are not visible
in the Hinode/SOT data. Extended filigree features are clearly fragmented into
separate BPs in the NST images. The majority of BPs have a circular shape,
however, some of them are elongated.

We found a weak positive proportional relationship between the lifetime, on one
hand, and the BP's size and intensity, on the other hand. So that brighter and
larger BPs tend to live longer.

The distribution function, PDF$(D)$, of the BP's size extents down to the
diffraction limit of NST (77 km). The saturation and the turnover of PDF$(D)$,
visible at smaller scales, might be caused by the fact that NST does not resolve
properly elements smaller than 77 km. This result is consistent with PDFs
derived from Hinode/G-band data (Utz et al. 2009): the Hinode PDFs also saturate
at the diffraction limit of the instrument.
The best approximation for the observed PDF$(D)$ is found to be a log-normal fit
with a mean value $m=4.157$ and a standard deviation $s=0.382$ of $ln(D)$. The
log-normal fit performs better than the exponential and power laws over the
entire range of diameters above 67 km. The mode of the log-normal distribution,
$e^m$, occurs at $D=64$ km, which is lower than the diffraction limit. Unlike
the power law, the log-normal distribution is not scale-free, therefore, there
is a limit on a minimal size of an elementary flux tube. However, the real
minimum size may not have been achieved yet from observations with modern
telescopes. The log-normal nature of the size distribution implies that the
fragmentation and merging processes are important mechanisms contributing into
evolution and dynamics of BPs (e.g., Abramenko \& Longcope 2005). Frequent
fragmentation and coalescence are clearly visible in the data set
movie\footnote{{\tt http://bbso.njit.edu/nst-galery.html}}.

The observed distribution function of the BP's lifetimes can be best fitted with
a log-normal approximation.
The NST lifetime distribution function qualitatively agrees with that reported
by de Wijn et al. (2005) inside the overlap interval, while Utz et al (2010)
reported an exponential distribution for the lifetime of Hinode BPs in the
interval of 2-12 minutes. In out study, the overall slope of the PDF$(LT)$ in
this particular 2-12 minutes interval is nearly the same as reported by Utz et
al. (2010), however, it appears that the power law is the best analytical fit
inside this particular interval.

About 98.6\%  of all BPs live less than 120 s (12 time steps). This remarkable
fact was not obvious from previous studies because an extremely high time
cadence was required. The fact indicates that the majority of BPs appear for
very short time (tens of seconds), similar to other transient features, for
example, chromospheric  rapid blue-shift events (RBEs), Rouppe van der Voort et
al. (2009). The most important point here is that these small and short living
BPs significantly increase dynamics (flux emergence, collapse into BPs, and
magnetic flux recycling) of unipolar network areas. These unipolar fields make
it difficult to invoke reconnection between the emerging and pre-existing flux
as an explanation for the RBEs. However, the fact that the network field is more
dynamic than expected may allow to apply the component reconnection approach.
The magnetic field is fragmented into flux concentrations with well defined
interfaces between them known as current sheets. Random motions of BPs, as well
as often appearance/disappearance of new flux concentrations, will change the
spatial distribution of the current sheets thus leading to component
reconnection and to release of energy.

It is interesting that we did not find a power law to be the best fit inside a
whole available scale interval for any of the tested parameters. Two reasons for
that might be advanced. First, a reign of the power law inside the interval
from zero to infinity is impossible unless the analyzed field is a
computer-generated monoflactal. Second, a power law implies
self-similarity - scaling with the same index - inside the entire range where
this law reigns. This means a monofractal structure of the field, and,
therefore, an absence of multifractality. At the same time, processes and
structures formed in a natural way possess different scaling indices for
different scale intervals; in other words, they are multifractals.
As broader range of scales becomes available for observations, the stronger is
the deviation of the observed distributions from a pure power law. 

This work was supported by NSF grants ATVI-0716512, ATM-0745744 and ATM0847126,
and NASA grants VNX08AJ20G, NNX08AQ89G, and NNX08BA22G, AFOSR (FA9550-09-1-0655)
and SOLAR-B subcontract through Lockheed (8100000779). Hinode is a Japanese
mission developed and launched by ISAS/JAXA, with NAOJ as domestic partner
and NASA and STFC (UK) as international partners. It is operated by these
agencies in co-operation with ESA and NSC (Norway).

\begin{table}[!ht]
\caption{\sf Parameters of the Distribution Functions for BP's Diameter, Lifetime and Intensity}
\footnotesize
\begin{center}
\begin{tabular}{lccc}
\hline
Parameter &  $D$, km &  $LT$, s & $I_{max}$, DN \\
\hline
\hline
Exponential &   &  & \\
Distribution&   &  & \\
\hline
 $c_1$    &   0.404$\pm$0.091   &   -3.20$\pm$0.29       &    6.18$\pm$0.31    \\
 $\beta$  &  -0.406$\pm$0.0008  &  -0.00427$\pm$0.00032  &  -0.00204$\pm$1e-05 \\
 $\chi^2$ &   0.26            &   1.11               &    3.04           \\
 $\Delta$ &  68 - 159         &   516 - 1226         &  4472 - 7540      \\
\hline
\hline
Power Law   &   &  & \\
Distribution&   &  & \\
\hline
$c_2$     &  15.50$\pm$1.18     &  3.05$\pm$0.18         &  43.2$\pm$1.3      \\
$\alpha$  &  -8.27$\pm$0.51     &  -1.99$\pm$0.07       &  -12.1$\pm$0.3     \\
$\chi^2$  &   0.60            &   0.06               &    0.63           \\
$\Delta$  &  160 - 254        &   91 - 706           &  4724 - 7804      \\
\hline
\hline
Log-Normal  &   &  & \\
Distribution&   &  & \\
\hline
$m$      &  4.157$\pm$0.034     &  4.64$\pm$0.12        &  8.29$\pm$0.01      \\
$s$      &  0.382$\pm$0.018     &  1.00$\pm$0.06        &  0.150$\pm$0.005    \\
$\chi^2$  &   0.030           &   0.071             &    18.0           \\
$\Delta$  &  $>$67            &   $>$100            &  $>$4200          \\
\hline
\hline
\end{tabular}
\end{center}

\end{table}

{}
\end{document}